\documentclass[letter]{aa}
\usepackage[utf8]{inputenc}
\usepackage{graphicx}
\usepackage{CJK}
\usepackage{txfonts}
\begin{document} 
\begin{CJK*}{UTF8}{gkai}

   \title{Measuring the spatio-temporal variation of fundamental physical constants using the OH sum rules}

   \author{Renzhi Su (苏仁智), 
          Minfeng Gu (顾敏峰)
          }

   \institute{Shanghai Astronomical Observatory, Chinese Academy of Sciences, 80 Nandan Road, Shanghai 200030, China\\
              \email{rzsu.astro@gmail.com}
             }

   \date{Received XX XX, 202X; accepted XX XX, 202X}

 
\abstract
   {In this paper, we introduce a method to constrain the variation of fundamental physical constants predicted by grand unified theories. This approach utilizes OH sum rules, which exhibit different forms depending on the physical conditions of the OH gas. Employing a least-squares technique, we can directly derive the relative velocity offsets among four OH lines in each ground state. This circumvents the need for spectral profile fitting, which is often unfeasible for spectra with complex profiles. Moreover, this method remains robust even when portions of the spectra are contaminated by radio frequency interference (RFI). Applying this method to a Galactic source G035.1+0.5, we obtained a stringent constraint of $\Delta X/X=(-1.8\pm5.1)\times10^{-6} $, where $X=\mu^{-1}\alpha^{2}g_{p}^{0.54}$. This method is expected to play a crucial role in extracting fundamental physics from large datasets of upcoming Square Kilometre Array (SKA) telescope.

}
   \titlerunning{Measuring variation of constants}
   \authorrunning{Renzhi Su et al.}
   
   \maketitle
   \nolinenumbers
%
\section{Introduction}
Over the past few decades, there has been considerable interest in measuring the spatio-temporal variations of fundamental constants, which are predicted to vary in extra-dimensional Kaluza-Klein and superstring theories \citep[e.g.][]{marciano1984,damour1994,li1998}. Throughout cosmic evolution, the contraction, expansion, or oscillation of extra dimensions could induce variations in the values of these constants. Detecting such variations, or establishing stringent upper limits, provides a crucial experimental test for current grand unified theories, in addition to constraining cosmological models \citep[e.g.][]{sandvik2002,thompson2013,thompson2013_2} and characterizing the dark energy equation of state \citep[][]{avelino2006,nunes2009}. However, because the amplitudes and timescales of these potential variations remain unknown, the experimental and observational efforts required to probe them face significant complexity and challenges. Of particular interest are the dimensionless constants: the fine-structure constant $\alpha \equiv e^2/(\hbar c)$ (expressed in c.g.s. system), the electron-proton mass ratio, $\mu \equiv m_e/m_p$, and the gyro-magnetic ratio of the proton (the proton g-factor), $g_{\rm p}$.

A critical and unavoidable step in previous studies aimed at measuring the variation of fundamental physical constants has been fitting observed spectra with a number of meaningful components. However, this approach is often unfeasible when spectra exhibit complex profiles or are contaminated by radio frequency interference (RFI). To circumvent these limitations, we propose a novel method based on OH sum rules to measure these variations.

The OH molecule is of particular interest in such studies because its transitions exhibit dependencies on $\alpha$, $\mu$, and $g_{\rm p}$, whereas transitions in most other molecules are sensitive to $\mu$. As given in \cite{chengalur2003,curran2004}, the sum frequency of OH 18-cm main lines $\nu_{1665}+\nu_{1667}$ is proportional to $\mu^{2.57}\alpha^{-1.14}R_{\infty}$, while the frequency difference of OH 18-cm satellite lines $\nu_{1720}-\nu_{1612}$ is proportional to $\mu^{0.72}\alpha^{2.56}g_{\rm p}R_{\infty}$, where $R_{\infty}$ is the Rydberg constant. Consequently, astronomical observations have demonstrated the utility of OH in probing potential variations in fundamental physical constants \citep[e.g.][]{gupta2018,kanekar2018,su2025a,su2025b,su2026}.

\section{The OH sum rule}\label{sec:ohrule}

Due to $\Lambda$-doubling and hyperfine structures, each rotational ground state of OH exhibits four energy sub-levels. For instance, the OH rotational ground state ($^{2}\Pi_{3/2}, J=3/2$) produces four hyperfine transition lines at rest-frame frequencies of $\nu_{1612}=1612.230825(15)$ MHz, $\nu_{1665}=1665.401803(12)$ MHz, $\nu_{1667}=1667.358996(4)$ MHz, and $\nu_{1720}=1720.529887(10)$ MHz \citep{hudson2006,lev2006}. The ratio of the number densities of the upper ($i$) and lower ($j$) energy levels is given by:
\begin{equation}\label{eq:number_distri}
\frac{N_{\rm i}}{N_{\rm j}}=\frac{g_{\rm i}}{g_{\rm j}}{\rm exp}\left(-\frac{h\nu}{kT_{\rm ij}^{\rm ex}}\right),
\end{equation}
where $g_{\rm i}$ and $g_{\rm j}$ are the statistical weights, and $T_{\rm ij}^{\rm ex}$ is the excitation temperature, which equals the gas kinetic temperature under local thermodynamic equilibrium (LTE) conditions. Based on the Equation \ref{eq:number_distri}, we obtain:
\begin{equation}\label{eq:freq_tex}
\frac{\nu_{1612}}{T_{1612}^{\rm ex}}+\frac{\nu_{1720}}{T_{1720}^{\rm ex}}=\frac{\nu_{1665}}{T_{1665}^{\rm ex}}+\frac{\nu_{1667}}{T_{1667}^{\rm ex}}.
\end{equation}

The optical depth is defined as:
\begin{equation}
\tau_{\nu}=\int \alpha_{\nu}d_{s},
\end{equation}
where $\alpha_{\nu}$ is the absorption coefficient and is expressed as: $\alpha_{\nu}=\frac{h\nu}{4\pi}(N_{\rm j}B_{\rm ji}-N_{\rm i}B_{\rm ij})\phi(\nu)$, where $B_{\rm ji}$ and $B_{\rm ij}$ are Einstein's B coefficients for absorption and stimulated emission,  and $\phi(\nu)$ is a normalized line profile function. Given $g_{\rm j}B_{\rm ji}=g_{\rm i}B_{\rm ij}$ and $A_{\rm ij}=\frac{2h\nu^{3}}{c^{2}}B_{\rm ij}$, where $A_{\rm ij}$ is Einstein's $A$-coefficient, the optical depth can be rewritten as:
\begin{equation}
\tau_{\nu}=\int \frac{c^{2}}{8\pi\nu^{2}} A_{\rm ij}N_{\rm i}(e^{h\nu/kT_{\rm ij}^{\rm ex}}  -1) \phi(\nu) d_{\rm s}.
\end{equation}

Assuming a homogeneous medium with $\int d_s=L$ and $kT_{\rm ij}^{\rm ex}\gg h\nu$, we have:

\begin{equation}\label{eq:vt_tau}
\frac{\nu}{T_{\rm ij}^{\rm ex}}=\frac{8\pi k\nu^{2}\tau_{\nu}}{Lc^{2}hA_{\rm ij}N_{\rm i}\phi(\nu)}.
\end{equation}
Taking this into Equation \ref{eq:freq_tex}, we obtain the following relation for the optical depths ($\tau$):
\begin{equation}\label{eq:OH18sumrule}
\tau_{1612}+\tau_{1720}=\tau_{1665}/5+\tau_{1667}/9
\end{equation}
\citep{robinson1967,tang2017}. This relationship among the optical depths is known as the `sum rule' and is valid across all velocity channels.  Note that the sum rule holds regardless of whether the gas is in LTE or non-LTE conditions. 

In cases where absorption occurs against continuum emission, the observed brightness temperature is given by $T_{\rm b} = (T_{\rm ex}-T_{\rm bg})(1-e^{-\tau})$, where $T_{\rm bg}$ is the brightness temperature of the background continuum source. Assuming LTE, optically thin, and a flat spectrum for the background continuum source, the observed line brightness temperatures satisfy:
\begin{equation}\label{eq:OH18sumrule_t}
T_{\rm b, 1612}+T_{\rm b, 1720}=T_{\rm b, 1665}/5+T_{\rm b, 1667}/9.
\end{equation}

Clearly, the relationships among the optical depths and brightness temperatures of the four OH 18 cm lines provide a robust means to measure the relative frequency (or velocity) shifts between these transitions, thereby probing potential variations in fundamental physical constants. 

Similarly, the sum rule exists for other OH rotational states, with the exception of the $^{2}\Pi_{1/2}, J=1/2$ state, which has only three transitions because the $F=0 \to 0$ transition is forbidden by selection rules. Using Equations \ref{eq:freq_tex} and \ref{eq:vt_tau}, we can derive the sum rule for each OH rotational state based on the Einstein's $A$-coefficients. For example, for the $^{2}\Pi_{3/2}, J=5/2$ state, the sum rule is:
\begin{equation}
\tau_{6017}+\tau_{6049}=\tau_{6031}/14+\tau_{6035}/20.
\end{equation}

For a comprehensive list of the Einstein's $A$-coefficients for various OH rotational states, see, e.g. \cite{destombes1977}.

\section{Methodology}

Considering four OH 18 cm lines, if fundamental physical constants vary, the standard sum rule (Equation \ref{eq:OH18sumrule}) must be modified to account for relative velocity shifts:
\begin{equation}\label{eq:OH18sumrule_vari}
\tau_{1612}(v) + \tau_{1720}(v + \Delta v_1) = \frac{1}{5}\tau_{1665}(v + \Delta v_2) + \frac{1}{9}\tau_{1667}(v + \Delta v_3),
\end{equation}
where $\Delta v_1$, $\Delta v_2$, and $\Delta v_3$ represent the velocity shifts of the 1720, 1665, and 1667 MHz transitions, respectively, relative to the 1612 MHz line. These shifts parameterize the potential variation in fundamental physical constants. The optimal values for $\Delta v_1$, $\Delta v_2$, and $\Delta v_3$ can be determined via least-squares fitting by minimizing the residual function $E$, defined as:
\begin{equation}\label{eq:least_squares}
\begin{split}
E = \sum_{v} [ \tau_{1612}(v) + \tau_{1720}(v + \Delta v_1) - \frac{1}{5}\tau_{1665}(v + \Delta v_2)\\
- \frac{1}{9}\tau_{1667}(v + \Delta v_3) ]^2.
\end{split}
\end{equation}

Next, we evaluate the applicability of the OH sum rules under various physical conditions, specifically addressing the thermodynamic state (LTE versus non-LTE), the nature of the transitions (absorption versus emission), and the optical depth (optically thin versus thick).

\begin{itemize}
    \item The gas is in LTE.\\
          1) Assuming the lines are observed in emission and there is no background radio continuum emission. The observed line brightness temperature is  $T_{\rm b} = T_{\rm ex}(1-e^{-\tau})$. Under LTE condition, the excitation temperature equals the gas kinetic temperature, $T_{\rm ex} = T_{\rm kin}$. If the gas is optically thin, then $T_{\rm b} = T_{\rm ex}(1-e^{-\tau})\approx T_{\rm kin}\tau$, yielding an optical depth of $\tau = T_{\rm b}/T_{\rm kin}$. Because $T_{\rm kin}$ is uniform across all transitions, it cancels out in every term of Equation \ref{eq:OH18sumrule}, meaning the OH sum rule can be used. In contrast, if the gas is optically thick, the optical depth must be expressed as $\tau = -\ln(1 - T_{\rm b}/T_{\rm kin})$. Due to this non-linear logarithmic dependence, $T_{\rm kin}$ no longer cancels out. Since $T_{\rm kin}$ is generally unknown, the OH sum rule cannot be applied in this situation.

          2) Assuming the lines are observed in absorption against background radio continuum emission. In such scenarios, the background brightness temperature typically far exceeds the excitation temperature ($T_{\rm bg} \gg T_{\rm ex}$). Therefore, $T_{\rm b} = -T_{\rm bg}(1-e^{-\tau})$. Because both $T_{\rm b}$ and $T_{\rm bg}$ are directly observable quantities, the optical depth $\tau$ can be explicitly derived regardless of whether the medium is optically thin or optically thick. Consequently, the OH sum rule is applicable. 

    \item The gas is in non-LTE. \\
    1) Assuming the lines are observed in emission and there is no background radio continuum emission. As the gas is in non-LTE, the $T_{\rm ex}$ is no longer uniform across different transitions.    Consequently, $T_{\rm ex}$ cannot be factored out of the terms in Equation \ref{eq:OH18sumrule}. As a result, the OH sum rule is applicable, regardless of whether the gas is optically thin or optically thick.
    
    2) Assuming the lines are observed in absorption against background radio continuum emission. Like the gas in LTE, the OH sum rule always can be used.
\end{itemize}

\section{Example}
\begin{figure}[h]
\centering
\includegraphics[width=0.45\textwidth]{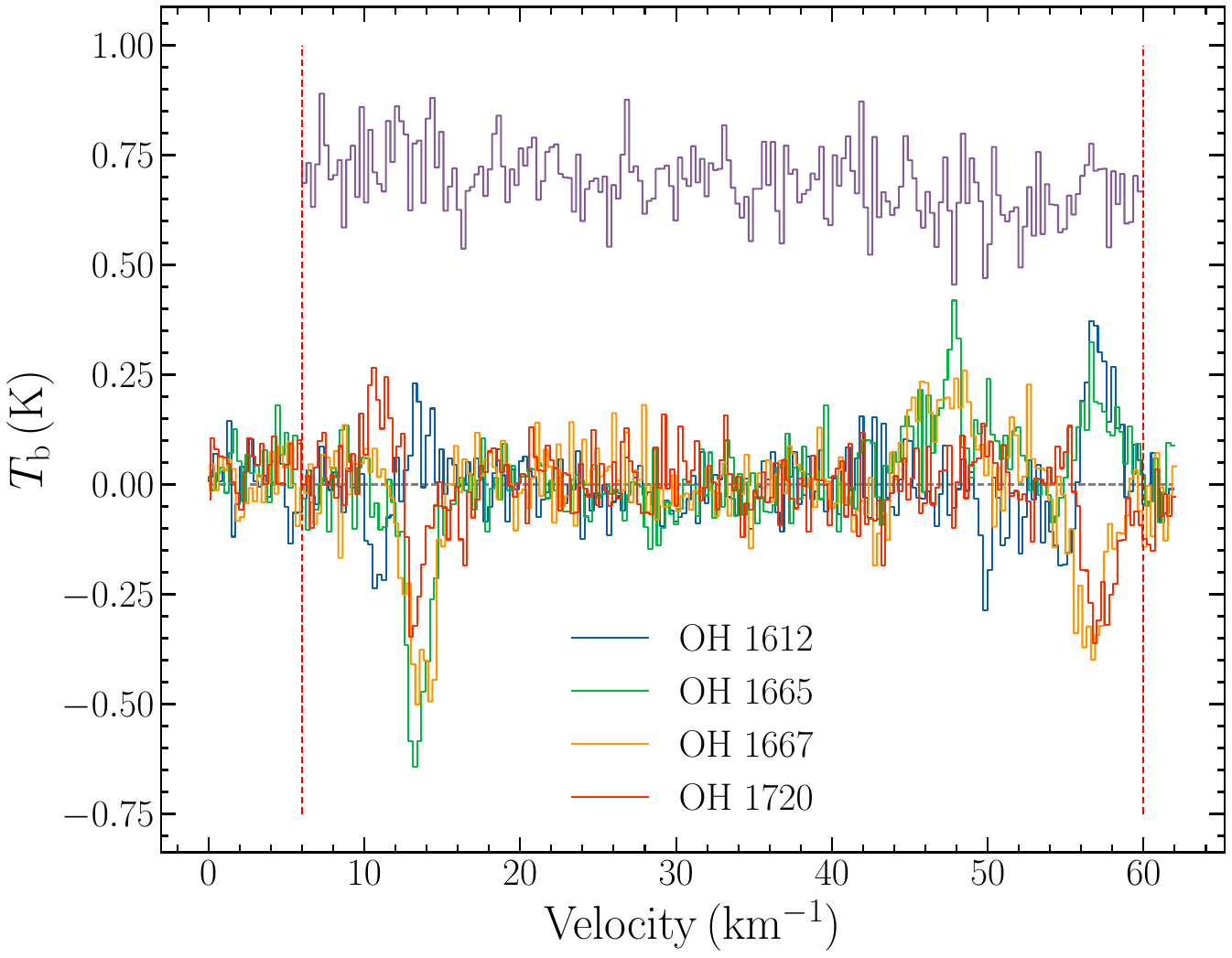}
\caption{Spectra of the OH 18 cm lines observed toward G035.1+0.5. The two vertical dashed lines mark the velocity boundaries of $6$ and $60~\mathrm{km\,s^{-1}}$, defining the range over which the least-squares fitting was performed to derive the velocity shifts $\Delta v_1$, $\Delta v_2$, and $\Delta v_3$. The purple curve represents $T_{1612}(v) + T_{1720}(v + \Delta v_1) - \frac{1}{5}T_{1665}(v + \Delta v_2)
- \frac{1}{9}T_{1667}(v + \Delta v_3)$, which has been vertically offset for visual clarity.
}\label{fig:ohlines}
\end{figure}

To demonstrate its practical applicability and advancement, we now apply the OH sum rule to observed OH 18 cm lines. We selected a Galactic source G035.1+0.5, toward which previous Arecibo observations detected all four OH 18 cm transitions \citep{tang2017}, as shown in Figure \ref{fig:ohlines}. The spectra clearly exhibit both emission and absorption features. Given the large beam size of the Arecibo telescope, it is reasonable to assume that the emission and absorption arise from physically distinct gas components along the line of sight. We further adopt the following assumptions: (1) the emitting gas is in LTE and is optically thin, then $\tau = T_{\rm b}/T_{\rm kin}$; and (2) the absorbing gas is also in LTE, is optically thin, and is observed against a background continuum source with a flat spectrum. Under these conditions, both the emission and absorption components independently satisfy Equation \ref{eq:OH18sumrule_t}. Consequently, their linear superposition (the total observed spectrum) satisfies Equation \ref{eq:OH18sumrule_t} as well. We note that the OH satellite lines could be conjugate (non-thermal). When pumping occurs, one satellite line appears in emission while the other one appears in absorption \citep[e.g.][]{van1995,darling2004,kanekar2018,su2026}.

\begin{figure*}[h]
\centering
\includegraphics[width=16cm]{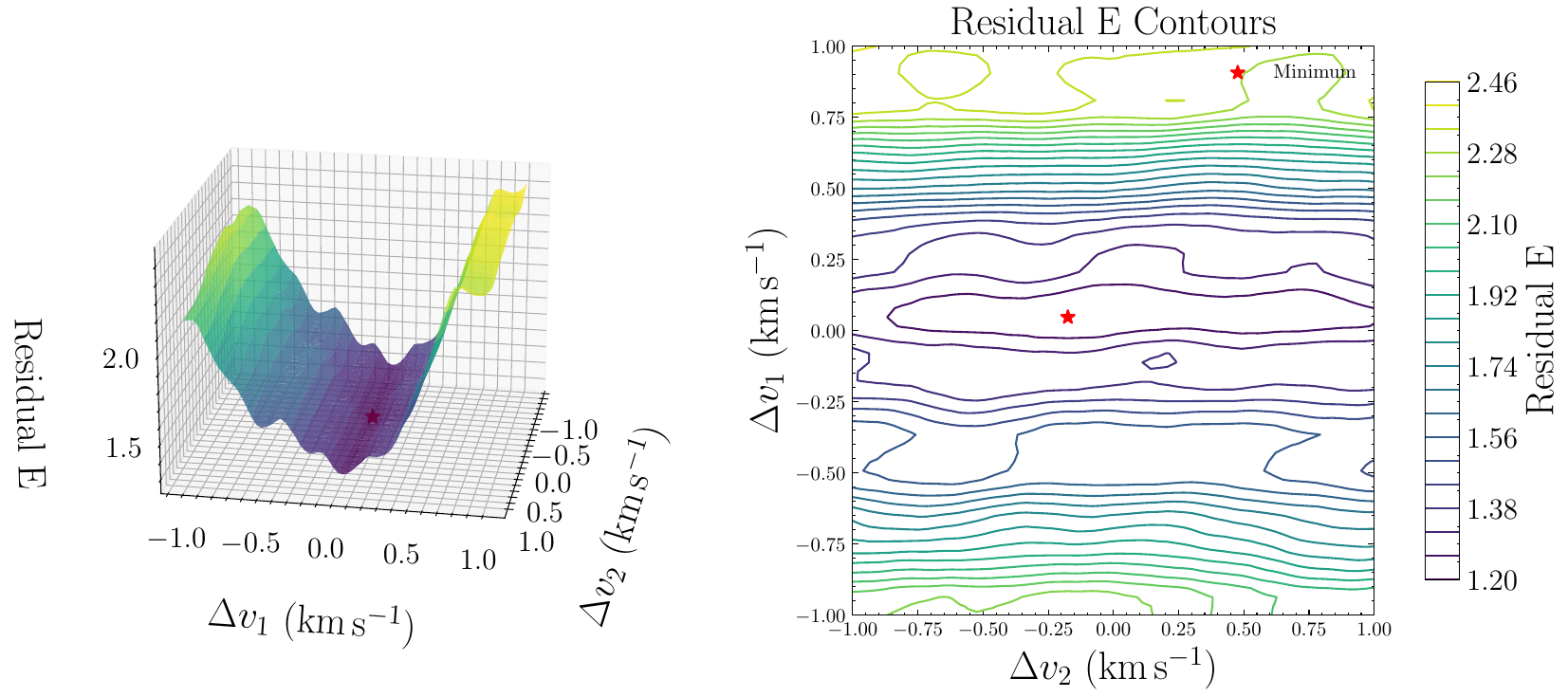}
\caption{The distribution of residual $E$ as a function of  $\Delta v_1$ and $\Delta v_2$. The red stars indicate the positions of $\Delta v_{1,\rm tmp}$ and $\Delta v_{2,\rm tmp}$ giving the preliminary minimum $E$.
}\label{fig:correlate_result}
\end{figure*}

We next employed a least-squares fitting procedure to determine the velocity shifts $\Delta v_1$, $\Delta v_2$, and $\Delta v_3$. In principle, the frequencies of OH 18 cm lines satisfy $\nu_{1612}+\nu_{1720}=\nu_{1665}+\nu_{1667}$; therefore, $\Delta v_3$ can be directly calculated from $\Delta v_1$ and $\Delta v_2$. Consequently, during the fitting process, we only treated $\Delta v_1$ and $\Delta v_2$ as free parameters, with $\Delta v_3$ being subsequently derived. To minimize the objective function $E$, we first sampled the parameter space of both $\Delta v_1$ and $\Delta v_2$ over a range of -1 to 1  $\mathrm{km\,s^{-1}}$ to find the preliminary optimal values, $\Delta v_{1,\rm tmp}$ and $\Delta v_{2,\rm tmp}$, which yielded a preliminary minimum $E$, as illustrated in Figure. \ref{fig:correlate_result}. Next, we utilized the differential evolution (DE) algorithm \citep{storn1997} as implemented in the \texttt{scipy.optimize.differential\_evolution } function from \textit{SciPy} \cite{virtanen2020} to obtain the optimal $\Delta v_{1,\rm opt}$ and $\Delta v_{2,\rm opt}$ by exploring a small parameter space centered on $\Delta v_{1,\rm tmp}$ and $\Delta v_{2,\rm tmp}$. 

The final best-fit values of $\Delta v_{1,\rm opt}$ and $\Delta v_{2,\rm opt}$, along with their associated uncertainties were derived through Monte Carlo simulations. By adding random noise characterized by the observed OH spectra to the underlying pure-signal models, we repeated the aforementioned fitting procedure 10,000 times. The means and standard deviations of the resulting distributions were then adopted as the final optimal values and their associated uncertainties. A key step involved above is the extraction of the pure-signal models from the observed spectra. A straightforward approach would be to fit the observed spectra with as many Gaussian components as required. Because the sole objective here is to obtain the pure-signal models, the specific Gaussian components used need not possess physical meaning. However, we adopted an alternative approach. We de-noised the spectra using a Fourier-based filtering approach: the fast Fourier transform (FFT) was applied to each spectrum, and only the low frequency components that accounted for a specified fraction (e.g., 90\%) of the total spectral power were retained. The de-noised signal, which serves as the pure-signal model, was then reconstructed via an inverse FFT. To validate this method, we tested it on simulated pure-signal models with injected noise. In our simulated dataset, the extracted pure-signal models contained slight residual noise compared to the input models; however, this residual noise was negligible compared to the initially injected noise. Therefore, the pure-signal models extracted from the observed OH spectra via the FFT method are sufficiently robust for our Monte Carlo simulations. Any residual noise present in these models is heavily dominated by the random noise injected during each iteration. Therefore, our final results remain unaffected by this extraction method.

The best-fit velocity shifts are $\Delta v_1 = -0.067 \pm 0.179~\mathrm{km\,s^{-1}}$, $\Delta v_2 = 0.400 \pm 0.642~\mathrm{km\,s^{-1}}$, from which we obtained $\Delta v_3 = -0.445\pm 0.667~\mathrm{km\,s^{-1}}$.

Systematic uncertainties may influence the results, originating from frequency calibration, laboratory rest frequencies, and intrinsic source variability.  Being phase-locked to a maser, radio spectrometers are very stable, rendering their associated uncertainty negligible. Furthermore, all OH observations were conducted within a six-month period \citep{tang2017}; thus, significant temporal evolution of the target source is not expected.  Laboratory rest frequencies are a key source of uncertainty. Given the frequency uncertainties in Section \ref{sec:ohrule}, the corresponding velocity uncertainties are $0.0028$, $0.0022$, $0.00072$, and $0.0017~\mathrm{km\,s^{-1}}$, respectively. Since these systematic uncertainties induced by the rest frequencies are substantially smaller than the statistical uncertainties of $\Delta v_1$, $\Delta v_2$, and $\Delta v_3$, they are considered negligible.

Given the frequency dependencies of OH lines on $\alpha$, $\mu$, and $g_{p}$ \citep{chengalur2003,curran2004}, we obtained the following constraints:
\begin{enumerate}
    \item by comparing the measured redshifts between $\nu_{1667}+\nu_{1665} $ and $\nu_{1667}-\nu_{1665} $, we have 
\begin{equation}\label{equ:compa_oh_main}
\frac{\Delta (\mu^{-1}\alpha^{2}g_{p}^{7.69})}{\mu^{-1}\alpha^{2}g_{p}^{7.69}} = (1.8\pm2.0)\times10^{-2};
\end{equation}
    
    \item by comparing the measured redshifts between $\nu_{1720}+\nu_{1612} $ and $\nu_{1720}-\nu_{1612} $, we have 
\begin{equation}\label{equ:compa_oh_sate}
\frac{\Delta (\mu^{-1}\alpha^{2}g_{p}^{0.54})}{\mu^{-1}\alpha^{2}g_{p}^{0.54}} = (-1.8\pm5.1)\times10^{-6}.
\end{equation}

\end{enumerate}

Combining Equation \ref{equ:compa_oh_main} and \ref{equ:compa_oh_sate}, we further obtained constraints:$\frac{\Delta g_{p}}{g_{p}} =(2.5\pm2.8)\times10^{-3}$ and $\frac{\Delta (\mu^{-1}\alpha^{2})}{\mu^{-1}\alpha^{2}} =(-1.4\pm1.5)\times10^{-3}$.

\section{Summary}

Probing the spatio-temporal variation of fundamental physical constants is a subject of profound interest in modern physics. Any robust detection would fundamentally transform our understanding of basic physics and the evolution of the Universe. Traditional methods for constraining these variations rely primarily on modeling and fitting observed spectral line profiles. However, spectral fitting becomes highly problematic or entirely unfeasible when the data are contaminated by radio frequency interference (RFI) or exhibit highly complex, blended line profiles. 

In this work, we introduce an approach based on the OH sum rules that circumvents these limitations. For the 18 cm transitions, the sum rule governing the optical depths is expressed in Equation \ref{eq:OH18sumrule}, which is always valid as long as the gas is thermal. Under specific assumptions, this relation translates directly to the observable brightness temperatures, as shown in Equation \ref{eq:OH18sumrule_t}. By applying a least-squares fitting technique to this relation, we can directly extract the relative velocity shifts among the four OH 18 cm lines, thereby providing  measurements or constraints on the variation of fundamental constants. 

Crucially, this method remains robust even when portions of the spectral baseline are masked due to RFI contamination. Furthermore, the approach does not require assumptions about the underlying line profile shapes. Consequently, a vast archive of OH spectra that have complex or blended profiles can now be effectively utilized. For example, the Southern Parkes Large-Area Survey in Hydroxyl (SPLASH) has yielded an abundance of such complex OH spectra \citep{dawson2022}, which are prime candidates for this new technique. 

Applying this method to a Galactic source G035.1+0.5, we successfully derived stringent constraints on the fractional variations of several combinations of fundamental constants: $\frac{\Delta (\mu^{-1}\alpha^{2}g_{p}^{0.54})}{\mu^{-1}\alpha^{2}g_{p}^{0.54}} = (-1.8\pm5.1)\times10^{-6}$, $\frac{\Delta (\mu^{-1}\alpha^{2})}{\mu^{-1}\alpha^{2}} = (-1.4\pm1.5)\times10^{-3}$, $\frac{\Delta g_{p}}{g_{p}} = (2.5\pm2.8)\times10^{-3}$, and $\frac{\Delta (\mu^{-1}\alpha^{2}g_{p}^{7.69})}{\mu^{-1}\alpha^{2}g_{p}^{7.69}} = (-1.8\pm2.0)\times10^{-2}$. 

Finally, this method is poised to play a pivotal role in the era of the Square Kilometre Array (SKA). With its unprecedented collecting area, the SKA will detect an enormous volume of OH spectra, and our sum-rule technique will be essential for extracting fundamental physics from these massive, complex datasets.

\begin{acknowledgements}
We thank Ningyu Tang who shared us the OH spectra of G035.1+0.5. MFG is supported by the National Science Foundation of China (grant 12473019), the Shanghai Pilot Program for Basic Research-Chinese Academy of Science, Shanghai Branch (JCYJ-SHFY-2021-013), the National SKA Program of China (Grant No. 2022SKA0120102), the National Science and Technology Major Project (2024ZD1100601), and the China Manned Space Project with No. CMS-CSST-2025-A07. RZS acknowledges the support from 
the National Science Foundation of China (grant 12603034), the Shanghai Super Postdoctoral Incentive Program (No. 2025322), the China Postdoctoral Science Foundation (Grant No. 2025M783232). and the National SKA Program of China (No. 2025SKA0130100).
\end{acknowledgements}

%
%
\bibliographystyle{aa} 
\bibliography{measuring_constants} 

@PREAMBLE{
 "\providecommand{\noopsort}[1]{}" 
 # "\providecommand{\singleletter}[1]{#1}%" 
}

@ARTICLE{marciano1984,
       author = {{Marciano}, W.~J.},
        title = "{Time variation of the fundamental 'constants' and Kaluza-Klein theories}",
      journal = {\prl},
         year = 1984,
        month = feb,
       volume = {52},
        pages = {489-491},
          doi = {10.1103/PhysRevLett.52.489},
       adsurl = {https://ui.adsabs.harvard.edu/abs/1984PhRvL..52..489M}
}

@ARTICLE{damour1994,
       author = {{Damour}, T. and {Polyakov}, A.~M.},
        title = "{The string dilation and a least coupling principle}",
      journal = {Nuclear Physics B},
         year = 1994,
        month = jul,
       volume = {423},
       number = {2-3},
        pages = {532-558},
          doi = {10.1016/0550-3213(94)90143-0},
archivePrefix = {arXiv},
       eprint = {hep-th/9401069},
 primaryClass = {hep-th},
       adsurl = {https://ui.adsabs.harvard.edu/abs/1994NuPhB.423..532D}
}

@ARTICLE{li1998,
       author = {{Li}, Li-Xin and {Gott}, J. Richard, III},
        title = "{Inflation in Kaluza-Klein theory: Relation between the fine-structure constant and the cosmological constant}",
      journal = {\prd},
         year = 1998,
        month = nov,
       volume = {58},
       number = {10},
          eid = {103513},
        pages = {103513},
          doi = {10.1103/PhysRevD.58.103513},
archivePrefix = {arXiv},
       eprint = {astro-ph/9804311},
 primaryClass = {astro-ph},
       adsurl = {https://ui.adsabs.harvard.edu/abs/1998PhRvD..58j3513L}
}

@ARTICLE{sandvik2002,
       author = {{Sandvik}, H{\r{a}}vard Bunes and {Barrow}, John D. and {Magueijo}, Jo{\~a}o},
        title = "{A Simple Cosmology with a Varying Fine Structure Constant}",
      journal = {\prl},
         year = 2002,
        month = jan,
       volume = {88},
       number = {3},
          eid = {031302},
        pages = {031302},
          doi = {10.1103/PhysRevLett.88.031302},
archivePrefix = {arXiv},
       eprint = {astro-ph/0107512},
 primaryClass = {astro-ph},
       adsurl = {https://ui.adsabs.harvard.edu/abs/2002PhRvL..88c1302B}
}

@ARTICLE{thompson2013,
       author = {{Thompson}, Rodger I. and {Martins}, C.~J.~A.~P. and {Vielzeuf}, P.~E.},
        title = "{Constraining cosmologies with fundamental constants - I. Quintessence and K-essence}",
      journal = {\mnras},
         year = 2013,
        month = jan,
       volume = {428},
       number = {3},
        pages = {2232-2240},
          doi = {10.1093/mnras/sts187},
archivePrefix = {arXiv},
       eprint = {1210.3031},
 primaryClass = {astro-ph.CO},
       adsurl = {https://ui.adsabs.harvard.edu/abs/2013MNRAS.428.2232T}
}

@ARTICLE{thompson2013_2,
       author = {{Thompson}, Rodger I.},
        title = "{A new substantive proton to electron mass ratio constraint on rolling scalar field cosmologies}",
      journal = {\mnras},
         year = 2013,
        month = may,
       volume = {431},
       number = {3},
        pages = {2576-2579},
          doi = {10.1093/mnras/stt355},
archivePrefix = {arXiv},
       eprint = {1302.5706},
 primaryClass = {astro-ph.CO},
       adsurl = {https://ui.adsabs.harvard.edu/abs/2013MNRAS.431.2576T}
}

@ARTICLE{avelino2006,
       author = {{Avelino}, P.~P. and {Martins}, C.~J.~A.~P. and {Nunes}, N.~J. and {Olive}, K.~A.},
        title = "{Reconstructing the dark energy equation of state with varying couplings}",
      journal = {\prd},
         year = 2006,
        month = oct,
       volume = {74},
       number = {8},
          eid = {083508},
        pages = {083508},
          doi = {10.1103/PhysRevD.74.083508},
archivePrefix = {arXiv},
       eprint = {astro-ph/0605690},
 primaryClass = {astro-ph},
       adsurl = {https://ui.adsabs.harvard.edu/abs/2006PhRvD..74h3508A}
}

@ARTICLE{nunes2009,
       author = {{Nunes}, N.~J. and {Dent}, T. and {Martins}, C.~J.~A.~P. and {Robbers}, G.},
        title = "{Reconstructing the evolution of dark energy with variations of fundamental parameters .}",
      journal = {\memsai},
         year = 2009,
        month = jan,
       volume = {80},
        pages = {785},
          doi = {10.48550/arXiv.0910.4935},
archivePrefix = {arXiv},
       eprint = {0910.4935},
 primaryClass = {astro-ph.CO},
       adsurl = {https://ui.adsabs.harvard.edu/abs/2009MmSAI..80..785N}
}

@ARTICLE{curran2004,
       author = {{Curran}, S.~J. and {Kanekar}, N. and {Darling}, J.~K.},
        title = "{Measuring changes in the fundamental constants with redshifted radio absorption lines}",
      journal = {\nar},
         year = 2004,
        month = dec,
       volume = {48},
       number = {11-12},
        pages = {1095-1105},
          doi = {10.1016/j.newar.2004.09.004},
archivePrefix = {arXiv},
       eprint = {astro-ph/0409141},
 primaryClass = {astro-ph},
       adsurl = {https://ui.adsabs.harvard.edu/abs/2004NewAR..48.1095C}
}

@ARTICLE{chengalur2003,
       author = {{Chengalur}, Jayaram N. and {Kanekar}, Nissim},
        title = "{Constraining the Variation of Fundamental Constants using 18cm OH Lines}",
      journal = {\prl},
         year = 2003,
        month = dec,
       volume = {91},
       number = {24},
          eid = {241302},
        pages = {241302},
          doi = {10.1103/PhysRevLett.91.241302},
archivePrefix = {arXiv},
       eprint = {astro-ph/0310764},
 primaryClass = {astro-ph},
       adsurl = {https://ui.adsabs.harvard.edu/abs/2003PhRvL..91x1302C}
}

@ARTICLE{kanekar2018,
       author = {{Kanekar}, Nissim and {Ghosh}, Tapasi and {Chengalur}, Jayaram N.},
        title = "{Stringent Constraints on Fundamental Constant Evolution Using Conjugate 18 cm Satellite OH Lines}",
      journal = {\prl},
         year = 2018,
        month = feb,
       volume = {120},
       number = {6},
          eid = {061302},
        pages = {061302},
          doi = {10.1103/PhysRevLett.120.061302},
archivePrefix = {arXiv},
       eprint = {1801.07688},
 primaryClass = {astro-ph.CO},
       adsurl = {https://ui.adsabs.harvard.edu/abs/2018PhRvL.120f1302K}
}

@ARTICLE{su2026,
       author = {{Su}, Renzhi and {Curran}, Stephen J. and {Darling}, Jeremy and {Gu}, Minfeng and {Aditya}, J.~N.~H.~S. and {Tang}, Ningyu and {Li}, Di and {Zheng}, Zheng},
        title = "{A stringent constraint on the fractional change of proton g-factor}",
      journal = {Science China Physics, Mechanics, and Astronomy},
         year = 2026,
        month = feb,
       volume = {69},
       number = {4},
          eid = {249511},
        pages = {249511},
          doi = {10.1007/s11433-025-2863-7},
archivePrefix = {arXiv},
       eprint = {2511.17998},
 primaryClass = {astro-ph.CO},
       adsurl = {https://ui.adsabs.harvard.edu/abs/2026SCPMA..6949511S}
}

@ARTICLE{su2025a,
       author = {{Su}, Renzhi and {An}, Tao and {Curran}, Stephen J. and {Busch}, Michael P. and {Gu}, Minfeng and {Li}, Di},
        title = "{Constraints on the Fractional Changes of the Fundamental Constants at a Look-back Time of 2.5 Myr}",
      journal = {\apjl},
         year = 2025,
        month = mar,
       volume = {981},
       number = {2},
          eid = {L25},
        pages = {L25},
          doi = {10.3847/2041-8213/adb843},
archivePrefix = {arXiv},
       eprint = {2502.14576},
 primaryClass = {astro-ph.GA},
       adsurl = {https://ui.adsabs.harvard.edu/abs/2025ApJ...981L..25S}
}

@ARTICLE{su2025b,
       author = {{Su}, Renzhi and {Curran}, Stephen J. and {Combes}, Fran{\c{c}}oise and {Gupta}, Neeraj and {Muller}, Sebastien and {Li}, Di and {Gu}, Minfeng},
        title = "{New constraints on the values of the fundamental constants at a look-back time of 7.3 Gyr}",
      journal = {\aap},
         year = 2025,
        month = jun,
       volume = {698},
          eid = {A154},
        pages = {A154},
          doi = {10.1051/0004-6361/202453407},
archivePrefix = {arXiv},
       eprint = {2505.07200},
 primaryClass = {astro-ph.CO},
       adsurl = {https://ui.adsabs.harvard.edu/abs/2025A&A...698A.154S}
}

@ARTICLE{gupta2018,
       author = {{Gupta}, N. and {Momjian}, E. and {Srianand}, R. and {Petitjean}, P. and {Noterdaeme}, P. and {Gyanchandani}, D. and {Sharma}, R. and {Kulkarni}, S.},
        title = "{Discovery of OH Absorption from a Galaxy at z {\ensuremath{\sim}} 0.05: Implications for Large Surveys with SKA Pathfinders}",
      journal = {\apjl},
         year = 2018,
        month = jun,
       volume = {860},
       number = {2},
          eid = {L22},
        pages = {L22},
          doi = {10.3847/2041-8213/aac9cd},
archivePrefix = {arXiv},
       eprint = {1806.00172},
 primaryClass = {astro-ph.GA},
       adsurl = {https://ui.adsabs.harvard.edu/abs/2018ApJ...860L..22G}
}

@ARTICLE{hudson2006,
       author = {{Hudson}, Eric R. and {Lewandowski}, H.~J. and {Sawyer}, Brian C. and {Ye}, Jun},
        title = "{Cold Molecule Spectroscopy for Constraining the Evolution of the Fine Structure Constant}",
      journal = {\prl},
         year = 2006,
        month = apr,
       volume = {96},
       number = {14},
          eid = {143004},
        pages = {143004},
          doi = {10.1103/PhysRevLett.96.143004},
archivePrefix = {arXiv},
       eprint = {physics/0601054},
 primaryClass = {physics.atom-ph},
       adsurl = {https://ui.adsabs.harvard.edu/abs/2006PhRvL..96n3004H}
}

@ARTICLE{lev2006,
       author = {{Lev}, Benjamin L. and {Meyer}, Edmund R. and {Hudson}, Eric R. and {Sawyer}, Brian C. and {Bohn}, John L. and {Ye}, Jun},
        title = "{OH hyperfine ground state: From precision measurement to molecular qubits}",
      journal = {\pra},
         year = 2006,
        month = dec,
       volume = {74},
       number = {6},
          eid = {061402},
        pages = {061402},
          doi = {10.1103/PhysRevA.74.061402},
archivePrefix = {arXiv},
       eprint = {physics/0608194},
 primaryClass = {physics.atom-ph},
       adsurl = {https://ui.adsabs.harvard.edu/abs/2006PhRvA..74f1402L}
}

@ARTICLE{robinson1967,
       author = {{Robinson}, B.~J. and {McGee}, R.~X.},
        title = "{OH Molecules in the Interstellar Medium}",
      journal = {\araa},
         year = 1967,
        month = jan,
       volume = {5},
        pages = {183},
          doi = {10.1146/annurev.aa.05.090167.001151},
       adsurl = {https://ui.adsabs.harvard.edu/abs/1967ARA&A...5..183R}
}

@ARTICLE{tang2017,
       author = {{Tang}, Ningyu and {Li}, Di and {Heiles}, Carl and {Yue}, Nannan and {Dawson}, J.~R. and {Goldsmith}, Paul F. and {Kr{\v{c}}o}, Marko and {McClure-Griffiths}, N.~M. and {Wang}, Shen and {Zuo}, Pei and {Pineda}, Jorge L. and {Wang}, Jun-Jie},
        title = "{OH Survey along Sightlines of Galactic Observations of Terahertz C+}",
      journal = {\apj},
         year = 2017,
        month = apr,
       volume = {839},
       number = {1},
          eid = {8},
        pages = {8},
          doi = {10.3847/1538-4357/aa67e9},
archivePrefix = {arXiv},
       eprint = {1703.05864},
 primaryClass = {astro-ph.GA},
       adsurl = {https://ui.adsabs.harvard.edu/abs/2017ApJ...839....8T}
}

@ARTICLE{destombes1977,
       author = {{Destombes}, J.~L. and {Marliere}, C. and {Baudry}, A. and {Brillet}, J.},
        title = "{The exact hyperfine structure and Einstein A-coefficients of OH: consequences in simple astrophysical models.}",
      journal = {\aap},
         year = 1977,
        month = aug,
       volume = {60},
       number = {1},
        pages = {55-60},
       adsurl = {https://ui.adsabs.harvard.edu/abs/1977A&A....60...55D}
}

@ARTICLE{dawson2022,
       author = {{Dawson}, J.~R. and {Jones}, P.~A. and {Purcell}, C. and {Walsh}, A.~J. and {Breen}, S.~L. and {Brown}, C. and {Carretti}, E. and {Cunningham}, M.~R. and {Dickey}, J.~M. and {Ellingsen}, S.~P. and {Gibson}, S.~J. and {G{\'o}mez}, J.~F. and {Green}, J.~A. and {Imai}, H. and {Krishnan}, V. and {Lo}, N. and {Lowe}, V. and {Marquarding}, M. and {McClure-Griffiths}, N.~M.},
        title = "{SPLASH: the Southern Parkes Large-Area Survey in Hydroxyl - data description and release}",
      journal = {\mnras},
         year = 2022,
        month = may,
       volume = {512},
       number = {3},
        pages = {3345-3364},
          doi = {10.1093/mnras/stac636},
archivePrefix = {arXiv},
       eprint = {2203.05131},
 primaryClass = {astro-ph.GA},
       adsurl = {https://ui.adsabs.harvard.edu/abs/2022MNRAS.512.3345D}
}

@ARTICLE{storn1997,
       author = {{Storn}, Rainer and {Price}, Kenneth},
        title = "{Differential Evolution - A Simple and Efficient Heuristic for global Optimization over Continuous Spaces}",
      journal = {Journal of Global Optimization},
         year = 1997,
        month = dec,
       volume = {11},
        pages = {341-359},
          doi = {10.1023/A:1008202821328},
       adsurl = {https://ui.adsabs.harvard.edu/abs/1997JGOpt..11..341S}
}

@ARTICLE{virtanen2020,
       author = {{Virtanen}, Pauli and {Gommers}, Ralf and {Oliphant}, Travis E. and {Haberland}, Matt and {Reddy}, Tyler and {Cournapeau}, David and {Burovski}, Evgeni and {Peterson}, Pearu and {Weckesser}, Warren and {Bright}, Jonathan and {van der Walt}, St{\'e}fan J. and {Brett}, Matthew and {Wilson}, Joshua and {Millman}, K. Jarrod and {Mayorov}, Nikolay and {Nelson}, Andrew R.~J. and {Jones}, Eric and {Kern}, Robert and {Larson}, Eric and {Carey}, C.~J. and {Polat}, {\.I}lhan and {Feng}, Yu and {Moore}, Eric W. and {VanderPlas}, Jake and {Laxalde}, Denis and {Perktold}, Josef and {Cimrman}, Robert and {Henriksen}, Ian and {Quintero}, E.~A. and {Harris}, Charles R. and {Archibald}, Anne M. and {Ribeiro}, Ant{\^o}nio H. and {Pedregosa}, Fabian and {van Mulbregt}, Paul and {SciPy 1.  0 Contributors}},
        title = "{SciPy 1.0: fundamental algorithms for scientific computing in Python}",
      journal = {Nature Medicine},
         year = 2020,
        month = feb,
       volume = {17},
        pages = {261-272},
          doi = {10.1038/s41592-019-0686-2},
archivePrefix = {arXiv},
       eprint = {1907.10121},
 primaryClass = {cs.MS},
       adsurl = {https://ui.adsabs.harvard.edu/abs/2020NaMet..17..261V}
}

@ARTICLE{van1995,
       author = {{van Langevelde}, Huib Jan and {van Dishoeck}, Ewine F. and {Sevenster}, Maartje N. and {Israel}, Frank P.},
        title = "{Anomalously Excited OH and Competition between Maser Transitions toward Centaurus A}",
      journal = {\apjl},
         year = 1995,
        month = aug,
       volume = {448},
        pages = {L123},
          doi = {10.1086/309613},
       adsurl = {https://ui.adsabs.harvard.edu/abs/1995ApJ...448L.123V}
}

@ARTICLE{darling2004,
       author = {{Darling}, Jeremy},
        title = "{A Laboratory for Constraining Cosmic Evolution of the Fine-Structure Constant: Conjugate 18 Centimeter OH Lines toward PKS 1413+135 at z = 0.24671}",
      journal = {\apj},
         year = 2004,
        month = sep,
       volume = {612},
       number = {1},
        pages = {58-63},
          doi = {10.1086/422450},
archivePrefix = {arXiv},
       eprint = {astro-ph/0405240},
 primaryClass = {astro-ph},
       adsurl = {https://ui.adsabs.harvard.edu/abs/2004ApJ...612...58D}
}
\end{CJK*}
\end{document}